\documentclass[aps,superscriptaddress,showpacs,showkeys,nofootinbib]{revtex4}
\usepackage{graphicx}

\newcommand{\eg}{e.g.,\,}
\newcommand{\ie}{i.e.,\,}
\newcommand{\be}{\begin{equation}}
\newcommand{\ee}{\end{equation}}
\newcommand{\bea}{\begin{eqnarray}}
\newcommand{\eea}{\end{eqnarray}}
\newcommand{\etal}{et al.}

\newcommand{\E}{{\cal E}}
\newcommand{\D}{{\cal D}}
\newcommand{\Order}{${\cal O}$}

\newcommand{\ra}{\rightarrow}

\newcommand{\lamzero}{{\lambdabar_0}}
\newcommand{\kapbar}{\overline{\kappa}}

\begin{document}

\bibliographystyle{apsrev}

\title{Including Absorption in  Gordon's Optical Metric}

\author{B. Chen}
\email{Bin.Chen-1@ou.edu}
\author{R. Kantowski}
\email{kantowski@nhn.ou.edu}
\affiliation{Homer L.~Dodge Department~of  Physics and Astronomy, University of
Oklahoma, 440 West Brooks, Room~100,  Norman, OK 73019, USA}
\date{\today}

\begin{abstract}
We show that Gordon's optical metric on a curved spacetime can be generalized 
to include absorption by allowing the metric to become complex. 
We demonstrate its use in the realm of geometrical optics by giving three 
simple examples. We use one of these examples to
compute corrected distance-redshift relations for 
Friedman-Lema\^itre-Robertson-Walker
models in which the cosmic fluid has an associated complex index of 
refraction that represents grey extinction.
We then fit this corrected 
Hubble curve to the type Ia supernovae data and provide a possible explanation 
(other than dark energy) of the deviation of these observations from dark matter predictions.  
\end{abstract}



\maketitle

\section{Introduction}\label{sec:intro}

Gordon \citep{Gordon} made the interesting observation that 
any solution to Maxwell's equations in a curved spacetime 
filled with a fluid whose electromagnetic properties can 
be described by a real permittivity $\epsilon(x)$ and a real
permeability $ \mu(x),$ or hence a real refraction index 
$n(x)=\sqrt{\epsilon\mu},$ can be found by solving a 
slightly modified version of Maxwell's equations 
in a related optical spacetime with vacuum values 
for the permittivity and permeability, \ie with 
$\epsilon(x)=1$ and $\mu(x)=1$, see Eq.\,(\ref{Max-inhomo-bar}). That is to say 
there is a one to one relation between solutions of Maxwell's equations
in these two spacetimes, the physical spacetime 
with an index of refraction and its physical metric, and the optical 
spacetime with $n=1$ and its optical metric. For traveling electromagnetic waves, 
the modified geometry of the optical metric accounts 
for the decrease of the wave speed which in physical spacetime is actually caused by 
an index of refraction $n>1$. 
Until now the open question was whether or not 
absorption could be incorporated into the optical metric.
In this 
paper we show that if the Maxwell field is 
a monochromatic wave the optical metric can 
be modified to account for absorption as well 
as refraction. That is, a solution to a slightly 
modified set of Maxwell's equations, without 
refraction or absorption, in the optical spacetime 
gives the  appropriately absorbed and refracted 
wave in physical spacetime. This conclusion extends to the superposition of 
multi-frequency waves 
as long as the optical properties are frequency independent. 

We assume that at some given frequency range the fluid's  electromagnetic 
properties are linear and isotropic relative to the fluid's unit 4-velocity $u_au^a=-1$, 
and can be summarized by a complex permittivity $\epsilon(x^a)$ and a complex permeability 
$\mu(x^a)$ defined on the four dimensional spacetime manifold.
Following \cite{Landau} we write a complex refraction index $N(x^a)$ as
\be
N=\sqrt{\epsilon\mu}\equiv n+i\kappa,
\label{N}
\ee 
where $n$ and $\kappa$ are respectively the real and imaginary parts.
The real electric field/magnetic induction bivector 
\footnote{${\rm Re}\{\cdot\}$ is the real part of the argument.} 
$F_{ab}(E,B)={\rm Re}\{F_{ab}^\lamzero\}$ 
 that represents a traveling  monochromatic wave has a geometrical optics  
expansion (see \cite{Ehlers2}) of the form
\be\label{WKB-1}
F_{ab}^\lamzero=
e^{iS/\lamzero}\left(A_{ab}+{\lamzero\over i}B_{ab}+O(\lamzero^2)\right), 
\ee 
and satisfies the homogeneous Maxwell equation\footnote{Square brackets $[\>\>]$ symbolize complete anti-symmetrization of the enclosed indices.}
\be
\partial_{[a}F_{bc]}^\lamzero=0,
\label{Max-homo}
\ee
 in both the physical and optical spacetimes. We designate the constant 
expansion parameter of geometrical optics by 
$\lamzero$ because we use its value to adjust the wavelength. It's positioning as a 
superscript or subscript is for convenience only. The real part of $S(x^a)$ is the 
usual eikonal function which determines the surfaces of constant phase for the wave. 
The $A_{ab}$ term represents the usual amplitude of the geometrical optics approximation and the $B_{ab}$ term is its first order 
correction (covariant components $F_{ab}, A_{ab}$, and $B_{ab}$ are identical in the physical and optical spacetimes). 
The constitutive relations for the contravariant components of the real
displacement/magnetic field bivector in physical spacetime $H^{ab}(D,H)={\rm Re}\{H^{ab}_\lamzero\}$ are given by
\be
H^{ab}_\lamzero={1\over\mu} \bar{F}^{ab}_\lamzero\equiv {1\over\mu}\bar{g}^{ac}\bar{g}^{bd}F_{cd}^\lamzero,
\label{H}
\ee
where the optical metric $\bar{g}_{ab}$ of \citet{Gordon} (which now becomes complex) has been used to 
raise the covariant indices of $F_{ab}^\lamzero$ to produce $\bar{F}^{ab}_\lamzero$. 
The complex optical metric is related 
to the real physical metric $g_{ab}=-u_au_b+g_{\perp\,ab}$ by
\be\label{Gordon-Metric}
\bar{g}_{ab}=(1-{1\over \epsilon\mu})u_au_b+g_{ab}=-\frac{1}{N^2}u_au_b+g_{\perp\,ab},
\ee 
with inverse
\be
\bar{g}^{ab}=(1-\epsilon\mu)u^au^b+g^{ab}=-N^2u^au^b+g_{\perp}^{ab}.
\label{Gordon-Metric-1}
\ee
The familiar source-free inhomogeneous Maxwell Equations in physical spacetime remain
\be
\nabla_bH^{ba}_\lamzero=0.
\label{Max-inhomo}
\ee
The more familiar form of the constitutive relations 
\bea\label{constitutive}
H^{ab}_\lamzero u_b&=&\epsilon F^{ab}_\lamzero u_b,\cr
F_{[ab}^\lamzero u_{c]}&=&\mu H_{[ab}^\lamzero u_{c]},
\eea
have been replaced in Eq.\,(\ref{H}) by an equivalent single equation using 
Gordon's metric.
Just as with a real optical metric (see \cite{Gordon, Ehlers1, Ehlers2, Kantowski}),  
the source-free inhomogeneous Maxwell equation (\ref{Max-inhomo}) can be 
rewritten as 
\be
\bar{\nabla}_b\left(\sqrt{\epsilon/\mu}\ \bar{F}^{ba}_\lamzero\right)=0.
\label{Max-inhomo-bar}
\ee
The covariant derivative in Eq.\,(\ref{Max-inhomo-bar}) is with respect to the complex optical metric, and except 
for the reciprocal of the impedance, $Z^{-1}=\sqrt{\epsilon/\mu}$\,, would be the same as Maxwell's vacuum inhomogeneous 
equations in the optical spacetime but without polarizable materials, \ie the components 
of $\bar{F}^{ba}_\lamzero$ were obtained in Eq.\,(\ref{H}) by simply raising 
indices on $F_{ab}^\lamzero$ using the optical metric.
Because $\epsilon$ and $\mu$ are ordinarily wavelength dependent, the values of  
 $N$ and $Z$  at a spacetime point depend on the particular geometrical optics wave being considered.

To proceed further with the geometrical optics approximation we must make assumptions about the size 
of the imaginary part, $\kappa$, of the index of refraction in Eq.\,(\ref{N}) 
at the frequency of interest. 
We will consider two types of cases; for the first type  
the imaginary part $\kappa$ is not small 
compared to the real part $n$ (at the wavelengths of interest) 
and for the second type it is, \ie $\kappa<<n$. 
The first type includes the absorption of low frequency waves in a conductor 
as well as the absorption of microwaves by water. 
The second includes a case of interest to us, the extinction of 
light waves traveling in a dilute gas.
In the first case the eikonal $S$ in Eq.\,(\ref{WKB-1}) has an imaginary 
part which is not negligible compared to the real part (see Sec.\,\ref{sec:Complex}) and  
in the second type, $S$ can be taken as real (see Sec.\,\ref{sec:Real}). In Sec.\,\ref{sec:Complex} 
we include two complex eikonal examples and in Sec.\,\ref{sec:Real} we give one real example 
which we then use to evaluate distance-redshift in standard cosmologies when absorption is present.
Because much of the algebra of the second type is included in 
the first we start with a complex $S$. 
We give some concluding remarks in Sec.\,\ref{sec:Discussion}.

\section{A Complex Eikonal}\label{sec:Complex}

In this section we develop the geometrical optics approximation for waves traveling in a medium where 
absorption on the scale of a wavelength cannot be neglected. Such significant absorption requires the use of a complex eikonal.
By inserting Eq.\,(\ref{WKB-1}) 
into Maxwell's Eqs.\,(\ref{Max-homo}) and (\ref{Max-inhomo-bar}) we find that 
$A_{ab}$ is of the form
\be
A_{ab}=-2k_{[a}\E_{b]},
\label{A}
\ee  
where $k_a\equiv \partial_aS$ is a complex null vector of the optical metric $(\bar{k}^ak_a=0)$ satisfying a 
complex geodesic like equation $\dot{\bar{k}}^a=0$ in the optical spacetime.
In general, the invariant derivative `\,$\dot{}$\,' is defined by
\be
`\,\dot{}\,\hbox{'}\equiv \bar{k}^b\bar{\nabla}_b,
\ee
and rather than being a directional derivative as it is when the metric is real, 
it becomes a complex partial differential operator.

Because we are interested in 
 ``homogeneous" waves, \ie those for which the surfaces 
of constant phase and constant amplitude coincide \cite{Landau}, we require that the spatial part of $\bar{k}^a$,
\ie the part of  $\bar{k}^a$ which is orthogonal to $u^a$,
be proportional to a real unit spacelike vector $\hat{k}^a$, $\hat{k}^b\hat{k}_b=1$. 
This direction defines the propagation direction for the wave as seen by the optical fluid.
As a consequence we can write 
\bea
\bar{k}^a&=& -(S_{,b}u^b)N(Nu^a+\hat{k}^a),\nonumber\\
k_a&=& -(S_{,b}u^b)(u_a+N\hat{k}_a).
\label{homogeneous}
\eea
The local period $T$ and decay time $T_d$ of the wave  
are related to changes in the real and imaginary parts of the eikonal 
as seen by an observer moving with the fluid, \ie by
\be
-(S_{,b}u^b)=\lamzero\left(\frac{2\pi}{cT}-i\frac {2}{cT_d}\right),
\label{T}
\ee
which are in turn related to the local wavelength $\lambda$ and absorption coefficient $\alpha$ through the complex index of refraction $N$ by
\be
\frac{2\pi}{\lambda}+i\frac{\alpha}{2}=N\left(\frac{2\pi}{cT}-i\frac {2}{cT_d}\right).
\label{lambda}
\ee
Maxwell's equations further restrict the electric field amplitude $ \E_a$ in Eq.\,(\ref{A}) 
by $\E_a\bar{k}^a=0,$ but leave the 
remaining freedom of definition $\E_a\ra \E_a+f(x)k_a$ (here $f(x)$ is an arbitrary complex function).
The first order (polarization dependent) correction to geometrical optics is given by
\be
B_{ab}=2(\E_{[a,b]}-k_{[a}\D_{b]}),
\label{B}
\ee
with a remaining freedom  $\D_a\ra \D_a+g(x)k_a$. Furthermore the propagation equation for the 
electric field amplitude $\bar{\E}^a$ is
\be
\dot{\bar{\E}}^a+\bar{\E}^a\theta+\bar{\E}^a\dot{\phi}={\bar{k}^a\over 2}
(\bar{\nabla}_b\bar{\E}^b+k_b\bar{\D}^b+2\phi_{,b}\bar{\E}^b),
\label{Edot}
\ee
where $2\phi$ is the natural logarithm of the reciprocal of the impedance, \ie $2\phi=\log{\sqrt{\epsilon/\mu}}$\,, and $\theta$ is defined as the divergence of $\bar{k}^a$. 
It is a generalization of the expansion rate of the ``null 
rays" defined by the complex vector field $\bar{k}^a,$  \ie 
\be
\theta\equiv{1\over 2}\bar{\nabla}_a\bar{k}^a=\frac{\dot{\sqrt{A}\ \ }}{\sqrt{A}},
\label{theta}
\ee
and for a real metric is conventionally interpreted \cite{Sachs} as  
the fractional rate of change of the observer independent cross-sectional area $A$ of a small beam of neighboring rays.
In what follows we choose a gauge where $\E_au^a=0$ which makes $\E_a$ spacelike and transverse to the 
wave's propagation direction, \ie $\E_a\hat{k}^a=0$.
By contracting Eq.\,(\ref{Edot}) with $\E_a$ 
we arrive at the propagation equation for the amplitude of plane polarized waves
\be
\dot{(\E_a\bar{\E}^a)}+2(\E_a\bar{\E}^a)(\theta+\dot{\phi})=0,
\ee
which can 
be simplified to read 
\be
\dot{ \left[(\E_a\bar{\E}^a)A \sqrt{\epsilon/\mu}\right] }=0.
\label{Eprop} 
\ee

The time averaged 4-flux seen by an observer moving with the optical fluid is in general

\be
S^a\equiv \frac{c}{8\pi}{\rm Re}\left\{\overstar{H}^{ac}F_{cb}-\frac{1}{4}\delta^a_{\>b} \overstar{H}^{dc}F_{cd}\right\}u^b,
\label{4P}
\ee
where   $\overstar{\{\cdot\}}$ stands for `the complex conjugate of'. 

When  $\bar{F}^{ab}$ in Eq.\,(\ref{H}) is restricted to the lowest order geometrical optics approximation,
Eq.\,(\ref{WKB-1}), and  is homogeneous, \ie satisfies Eq.(\ref{homogeneous}),
we have  
\be
S^a=\frac{c}{8\pi}e^{-2S_I/\lamzero}(\E_a\overstar{\bar{\E}}^a)|S_{,b}u^b|^2{\rm Re}\left\{\sqrt{\epsilon/\mu}\right\}
\left[{\rm Re}{\{N\}}\,u^a+\hat{k}^a\right],
\label{4S}
\ee 
where $S_I$ is the imaginary part of the eikonal $S$, and ${\rm Re}{\{N\}}$ is the usual index of refraction, see Eq.\,(\ref{N}). 
The coefficient of the fluid velocity $u^a$ in $S^a$ is the time average of the energy density ($\times c$) and the coefficient of
$\hat{k}^a$ is the time average of the magnitude of the Poynting vector, 
both measured by observers moving with the optical fluid. 
Equation (\ref{4S}) shows that energy in the single frequency geometrical optics wave is transferred in the $\hat{k}^a$ direction with a speed of $c/n$ by this wave. In the next two subsections we give two concrete examples where $S$ is complex. 

\subsection{Plane Waves in Minkowski Spacetime}

To make contact with familiar examples in classical electrodynamics, 
we start with a plane wave propagating in an optical fluid which is 
at rest in flat spacetime. We assume $\epsilon$ and $\mu$ have only 
a z-dependence, and study waves propagating along the $z$ direction, 
starting at $z=-\infty$. To suppress reflections and to make the 
geometrical optics approximation valid, we assume that $\epsilon$ anf $\mu$ vary 
slowly over a wavelength  $\lambda$, \ie  
$\epsilon_{,z}\lambda\ll 1,$ $\mu_{,z}\lambda\ll 1$.
 The physical metric is flat Minkowskian, the fluid's 4-velocity is $u^a=\delta^a_0$, 
and the optical metric, Eq.\,(\ref{Gordon-Metric}), is
\be
\bar{ds}^2=-\frac{(cdt)^2}{N(z)^2}+dx^2+dy^2+dz^2.
\ee 
From Eq.\,(\ref{homogeneous}) we find the complex wave vector 
\bea
\bar{k}^a&=&N(N, 0,0, 1),\nonumber\\
k_a&=&(-1, 0,0, N),
\eea 
with the complex eikonal
\be
S\equiv S_R+i S_I=\left[-ct+\int_{-\infty}^{z}{n(z')dz'}\right]+i\,\left[\int_{-\infty}^{z}{\kappa(z')dz'}\right].
\ee 
Equations (\ref{T}) and (\ref{lambda}) reduce to 
\bea
1&=&\lamzero\left(\frac{2\pi}{cT}\right),\nonumber\\
\frac{2\pi}{\lambda}+i\frac{\alpha}{2}&=& N\left(\frac{2\pi}{cT}\right),
\eea
which give the wave a constant frequency 
$\nu\equiv1/T=c/(2\pi\lamzero),$
but a $z$ dependent wavelength 
$\lambda=2\pi\lamzero/n(z)$
and  a $z$ dependent absorption coefficient
$\alpha=2\kappa(z)/\lamzero.$
For this wave we have chosen the scale of $S$ so that the geometrical optics expansion parameter $\lamzero$ is the wave's 
rationalized wavelength in the absence of refractive material. 
This wave corresponds to a constant and uniform source (at $z=-\infty$) 
which resulted in $T_d=\infty$ in Eqs.\,(\ref{T}) and (\ref{lambda}).

The expansion defined in Eq.\,(\ref{theta}) vanishes for this plane wave, \ie $\theta=0.$ If it is linearly polarized along the $x$ direction the  amplitude of the 
 field 
is $(0,\E^x,0,0)$.  
The propagation equation (\ref{Eprop}) then simplifies to 
\be\label{conservation}
\dot{[\E^x({\epsilon}/{\mu})^{1/4} ]}=0
\ee which implies 
\be
\E^x\,\left(\frac{\epsilon}{\mu}\right)^{1/4} =f\left(-ct+\int_{-\infty}^{z}{N(z')dz'}\right),
\label{Ex}
\ee 
where the function $f$ reflects the time dependence of the source amplitude $\E(t,z)$ at the source, \ie at $z= -\infty$. For a 
stable plane wave source, we simply put $f={\rm constant}$.

To compute the energy flux we can use either the spatial part of the general  result Eq.\,(\ref{4S}),  
or because the physical metric is flat and the fluid is at rest,
use the familiar 3-D electric and magnetic fields from Eqs.\,(\ref{WKB-1}) and (\ref{A})
\bea
{\bf E}&=&e^{iS/\lamzero}{\E}^x\,\hat{\hbox{{\bf \i}}},\nonumber\\
{\bf H}&=&\sqrt{\epsilon/\mu}\ \hat{\bf k}\times{\bf E},
\eea
to evaluate the time averaged Poynting vector directly
\bea\label{s_vec}
{\bf S}&=&\frac{c}{8\pi}{\rm Re}\{{\bf E}\times{\overstar{\bf H}}\},\cr
&=&\frac{c}{8\pi}{\rm Re}\{\sqrt{\epsilon/\mu}\}|\E^x|^2e^{-2S_I/\lamzero}\hat{\bf k}.
\eea 
The magnitude of ${\bf S}$ can be evaluated using Eq.\,(\ref{Ex}) as
\be
S(z)=\frac{\cos\beta(z)}{\cos\beta(-\infty)}e^{-2S_I(z)/\lamzero}S(-\infty),
\label{S-plane}\ee
where $S(-\infty)$ is the flux at the source
and where
\be
\cos \beta\equiv \frac{ {\rm Re}\{\sqrt{\epsilon/\mu}\}} {\sqrt{|\epsilon/\mu|}}=\frac{{\rm Re}\{Z\}}{|Z|}.
\ee 
The phase $\beta$ represents the angle by which ${\bf H}$ field lags behind the ${\bf E}$ field, 
and $\cos\beta$ is the familiar power factor in the language of circuit analysis. 
From the imaginary part of the eikonal we can now easily write down the relation 
between the absorption coefficient $\alpha$  and the classical optical depth 
$\tau$ \citep{Mihalas}
\be \tau(z)=2\frac{\omega}{c}\int^{z}_{-\infty}{\kappa(z')dz'}=\int^{z}_{-\infty}{\alpha(z')dz'}.
\label{tau-flat}
\ee

\subsection{Spherical Waves in a Static Spherically Symmetric Spacetime}

Because this case is similar to the above, we truncate the discussion and give mainly the results. 
For an isotropic monochromatic source at rest at the origin of a static spherically symmetric spacetime 
which is emitting radiation 
at a steady rate into an optical fluid, which is also at rest, and whose optical properties depend only on the 
distance from the origin, we have an optical metric of the form
\be
d\bar{s}^2=-\frac{e^{2\Phi(r)}}{N(r)^2}(cdt)^2+e^{2\Psi(r)}dr^2+r^2(d\theta^2+\sin^2\theta d\phi^2),
\ee
and a fluid at rest with respect to the non-rotating Killing flow, \ie $u^a=e^{-\Phi}\delta^a_0$. 
For the radial null vectors in Eq.\,(\ref{homogeneous}) we have,
\bea
\bar{k}^a&=&N\left(Ne^{-2\Phi}, e^{-\Phi-\Psi},0,0\right),\cr
 k_a&=&(-1,Ne^{-\Phi+\Psi},0,0).
\eea
The complex eikonal is 
\be
S=\left[-ct+\int_{0}^{r}{n(r')e^{-\Phi+\Psi}dr'}\right]+i\,\left[\int_{0}^{r}{\kappa(r')e^{-\Phi+\Psi}dr'}\right].
\ee 
Equations (\ref{T}) and (\ref{lambda}) simplify to
\bea
e^{-\Phi}&=&\lamzero\left(\frac{2\pi}{cT}\right),\nonumber\\
\frac{2\pi}{\lambda}+i\frac{\alpha}{2}&=& N\left(\frac{2\pi}{cT}\right),
\eea
and  give an $r$ dependent 
frequency 
$\nu\equiv1/T=c\,e^{-\Phi}/(2\pi\lamzero),$
an $r$  dependent  wavelength 
$\lambda=2\pi\lamzero e^{\Phi}/n(r)$ 
and an $r$ dependent absorption coefficient
$\alpha=2\kappa(r)e^{-\Phi}/\lamzero.$ 
The expansion parameter $\theta$ of Eq.\,(\ref{theta}) is
\be
\theta= N\frac{e^{-\Phi-\Psi}}{r}=\frac{\dot{r}}{r}=\frac{\dot{\sqrt{A}}}{\sqrt{A}}.
\ee

For a time independent source Eq.\,(\ref{Eprop}) now  gives 
\be
\dot{\left[r(\epsilon/\mu)^{1/4}\E\right]}=0\,,
\label{transport-solution}
\ee
where the polarization vector has been written in the form $\E^a=(0,0,\E/r,0)$ and points in the $\hat{\bf e}_\theta$ direction.
The flux measured by $u^a=e^{-\Phi}\delta^a_0$ is found from Eq.\,(\ref{4S}) to be
\bea
S(r)&=&\frac{c}{8\pi}e^{-2S_I/\lamzero}|\E|^2(u^ak_a)^2{\rm Re}\{\sqrt{\epsilon/\mu}\},\cr
&=& e^{-\tau(r)}\frac{\cos\beta(r)}{\cos\beta(0)} e^{-2\Phi(r)+2\Phi(0)}\frac{{\cal L}}{4\pi r^2},
\eea
where ${\cal L}$ is the total isotropic power radiated by the stationary point source in a narrow frequency range. The 
optical depth $\tau$ changes from Eq.\,(\ref{tau-flat}) to
\bea
 \tau(r)&=&2\lamzero^{-1}\int^r_0{\kappa(r')e^{-\Phi+\Psi}dr'},\\
&=&\int^r_0{\alpha(r')e^{\Psi}dr'}.
\eea
The spherical result differs from the plane wave result of Eq.\,(\ref{S-plane})  by a decrease of the flux 
caused by the wave's expansion ($A^{-1}\propto r^{-2}$) and by a frequency shift in the 
wave as it moves through the changing gravity field.

\section{A Real Eikonal}\label{sec:Real}

Waves traveling in spacetimes where the imaginary part of the index of refraction is much smaller 
than the real part  must still satisfy the same set of Maxwell's equations (\ref{Max-homo}) and 
(\ref{Max-inhomo-bar}) as before but for them the eikonal $S$ in Eq.\,(\ref{WKB-1}) can be taken as real. 
The complex index of refraction is caused by a complex permittivity 
$\epsilon=\epsilon_R+i\epsilon_I$ and/or a complex permeability $\mu=\mu_R+i\mu_I$ 
whose imaginary parts are much smaller than their real parts. Consequently we  write $N$ as  
\be
N=n+i\kappa=n\left(1+i\lamzero\kapbar+{\cal O}(\lamzero^2)\right),
\label{tilde-N}
\ee
where $n\equiv \sqrt{\epsilon_R\mu_R}$ and the constant parameter $\lamzero$ is 
the same parameter used to keep track of the various orders in the geometrical optics expansion. 
The  absorption coefficient $\alpha$  of Eq.\,(\ref{lambda}) is related to $\kapbar$ 
by  $\alpha=2\kapbar\lamzero/\lambdabar$. 
The optical metric components of Eqs.\,(\ref{Gordon-Metric}) and (\ref{Gordon-Metric-1}) are
\bea
\bar{g}_{ab}&=&-\frac{1}{N^2}u_au_b+g_{\perp\,ab}= \tilde{g}_{ab}-2\frac{\lamzero}{i}\frac{\kapbar}{n^2}u_au_b+{\cal O}(\lamzero^2),\\
\bar{g}^{ab}&=&-N^2u^au^b+g_{\perp}^{ab}= \tilde{g}^{ab}+2\frac{\lamzero}{i}\kapbar n^2u^au^b+{\cal O}(\lamzero^2),
\eea
where the \Order($\lamzero^0$) term of the optical metric, $\tilde{g}_{ab}$,  and its inverse, $\tilde{g}^{ab}$, are real  
\bea\label{tilde-g}
\tilde{g}_{ab}&=&-\frac{1}{n^2}u_au_b+g_{\perp\,ab},\\
\tilde{g}^{ab}&=&-n^2u^au^b+g_{\perp}^{ab}.
\eea
When the geometrical optics expansion of Eq.\,(\ref{WKB-1}) is inserted into Eqs.\,(\ref{Max-homo}) and (\ref{Max-inhomo-bar})
 the  \Order($\lamzero^{-1}$) terms result in Eq.\,(\ref{A}) again, but now with $k_a$ 
the gradient of the real eikonal $S$ 
and null with respect to the \Order($\lamzero^0$)  optical metric $\tilde{g}_{ab}$, \ie 
\bea
\tilde{k}^ak_a&=&0,\\
\dot{\tilde{k}}^a=\frac{d\tilde{k}^a}{D\ell}\,&\equiv&\tilde{k}^b\tilde{\nabla}_bk^a=0.
\eea
This gives real geodesics $x^a(\ell)$ with real tangents 
$dx^a/d\ell=\tilde{k}^a\equiv\tilde{g}^{ab}k_b$, and makes the real metric $\tilde{g}_{ab}$ the important geometric quantity rather 
than the complex $\bar{g}_{ab}$.
The ` $\dot{}$ ' derivative is now the familiar derivative with respect to an affine parameter $\ell$ along null geodesics. 
Equations (\ref{homogeneous}) are still valid except $N$ is replaced by its real part $n$ and complex $\bar{k}^a$ by the real $\tilde{k}^a$.
Equations (\ref{T}) and (\ref{lambda}) for the frequency and wavelength are replaced by 
\be
-(S_{,b}u^b)=\lamzero\left(\frac{2\pi}{cT}\right),
\label{real-T}
\ee
and
\be
\frac{2\pi}{\lambda}=n\left(\frac{2\pi}{cT}\right).
\label{real-lambda}
\ee
The polarization vector $\E_a$ in Eq.\,(\ref{A}) is now constrained to be orthogonal to $\tilde{k}^a$ 
and can again be chosen orthogonal to the fluid $u_a$. The \Order($\lamzero^1$) correction terms $B_{ab}$ are 
still given by Eq.\,(\ref{B}), the expansion $\theta$ in Eq.\,(\ref{theta})  is computed using $\tilde{k}^a$,
 and the covariant derivatives are all with respect to the real $\tilde{g}_{ab}$ Christoffel connection. 
The only equation that contains a new term, the extinction term,  is the propagation equation for the 
amplitude $\E^a$ that replaces Eq.\,(\ref{Edot}),
\be
\dot{\tilde{\E}}^a+\tilde{\E}^a\theta+\tilde{\E}^a\dot{\phi}+\kapbar n^2(u^dk_d)^2\tilde{\E}^a={\tilde{k}^a\over 2}
(\tilde{\nabla}_b\tilde{\E}^b+k_b\tilde{\D}^b+2\phi_{,b}\tilde{\E}^b).
\label{tilde-Edot}
\ee
The phase $\phi$ is now computed similarly as for Eq.\,(\ref{Edot}) but now using only the 
\Order($\lamzero^0$) terms  of the impedance  
$2\phi=\log\sqrt{\epsilon_R/\mu_R}$. 
The integral of Eq.\,(\ref{tilde-Edot}) which replaces Eq.\,(\ref{Eprop}) now contains an affine parameter integral 
\be
\log{\left[(\overstar{\E}_a\tilde{\E}^a)A \sqrt{\epsilon_R/\mu_R}\right]} =-2\int \,\kapbar n^2(u^dk_d)^2d\,\ell\,.
\label{tilde-Eprop} 
\ee 
When the time averaged Poynting vector, Eq.\,(\ref{4P}), is evaluated Eq.\,(\ref{4S}) is replaced  by
\be
S^a=\frac{c}{8\pi}(\E_a\overstar{\tilde{\E}}^a)(S_{,b}u^b)^2\left(\sqrt{\epsilon_R/\mu_R}\right)
\left[n\,u^a+\hat{k}^a\right].
\label{tilde-4S}
\ee 

Comparing Eqs.\,(\ref{4S}) with (\ref{tilde-4S}) the effects of absorption can easily be seen to 
have shifted from the eikonal where it belongs if absorption is significant over wavelength scales, 
to the slowly changing amplitude $\E_a$  where it belongs if extinction is significant only over many wavelengths.  

In the next section we give an example of absorption with a real eikonal from observational cosmology. 
We evaluate luminosity-distance in a Friedman-Lema\^itre-Robertson-Walker (FLRW) universe when extinction occurs.

\subsection{Robertson-Walker Spacetime} 

As an example of weak absorption we apply our complex extension of Gordon's optical 
theory to a Robertson-Walker (RW) universe filled with
a time dependent index of refraction $N(t)=n(t)+i\kappa(t)$ [see Eq.\,(\ref{tilde-N})] and obtain for the real optical metric, Eq.\,(\ref{tilde-g}), 
\be
\tilde{ds}^2=-\frac{(cdt)^2}{n(t)^2}+R^2(t)\left\{{dr^2\over 1-kr^2}+r^2(d\theta^2+\sin^2\theta\, d\phi^2)\right\}.
\label{tilde-FRW}
\ee 
The familiar curvature parameter $k=(1,0,-1)$ distinguishes respectively between  spatially closed, flat, and open models. 
The radial outgoing null geodesics can be solved immediately to give null vectors 
\bea\label{RW-k}
\tilde{k}^a&=&R_0\left({n\over R},{\sqrt{1-kr^2}\over R^2},0,0\right),\\
k_a&=&R_0\left(-{1\over nR},{1\over \sqrt{1-kr^2}},0,0\right),
\eea  
and the related real spherically symmetric eikonal centered at the emission point, $t=t_e,\, r=0$,
\be
S(t,r)=S_R=R_0\left(-\int^{t}_{t_e}{c\,dt\over n(t)R(t)}+{\rm sinn}^{-1}[r]\,\right),
\ee 
where
\be
{\rm sinn}[r] \equiv \cases{\sin[r] & $k=+1,$\cr r & $k=0,$\cr \sinh[r] & $k=-1$.}
\ee 
The constant $R_0$ is the current radius of the universe [from Eq.\,(\ref{tilde-FRW})] and has been introduced so 
that the geometrical optics expansion parameter $\lamzero$ corresponds to the rationalized wavelength 
of the wave when it reaches an observer at $t_0,r_0$ from an emitting source at $t_e,r=0$.
From Eqs.\,(\ref{real-T}) and (\ref{real-lambda}) we have the wavelength and frequency redshifts
\bea
1+z&\equiv&\frac{\lambda_0}{\lambda}=\frac{R_0}{R},\cr
1+z_n&\equiv&\frac{\nu}{\nu_0}=\frac{n_0R_0}{n\>R}.
\eea 
which are thus related by
\be
(1+z_n)=\frac{n_0}{n\,}(1+z).
\label{zn}
\ee
The conventional RW redshift $(1+z)\equiv R_0/R$ is valid for both wavelength and frequency 
when the refracting and absorbing material is absent, \ie when $N=1$.
We see that even with refraction and absorption the wavelength redshift remains as in RW cosmology. 
The frequency redshift, 
however, is affected by the real part of the index of refraction, $n$, but not by the imaginary part, $\kappa$,
\ie not by extinction.  

To evaluate the apparent brightness of a source we need to evaluate the magnitude of the spatial 
part of the Poynting vector, Eq.\,(\ref{tilde-4S}), which requires that we know the area $A$ in Eq.\,(\ref{tilde-Eprop}). 
For  a spherical wave emanating from the comoving origin, $A\propto (rR)^2$, 
which is confirmed  by evaluating the expansion parameter $\theta$ using Eq.\,(\ref{RW-k}). 
From the transport Eq.\,(\ref{tilde-Eprop}) we have 
\be
(\overstar{\E}_a\tilde{\E}^a)\sqrt{\epsilon_R/\mu_R} 
\propto\frac{e^{-\tau}}{(Rr)^2},
\label{RW-Eprop}
\ee 
where the optical depth $\tau(t)$ from $t$ to  $t_0$ is 
\bea
\tau(t) &=& 2c\int^{t_0}_t \,\frac{\kapbar(t')}{\,n(t')}\frac{R_0}{R(t')}\,dt'\,,\\
&=&c\int^{t_0}_t \,\frac{\alpha(t')}{\,n(t')}\,dt'\,.
\eea

When looking back in time from the observer, Eq.\,(\ref{tilde-4S}) gives the 
 magnitude of the spatial part of flux 
\be
S(t)\propto \frac{e^{-\tau(t)}}{[n(t)R(t)]^2[r(t)R(t)]^2}.
\ee 
 The apparent luminosity $L$ in a narrow band is just the value of $S(t)$ at the observer, $t=t_0$. 
If the absorption is frequency independent in the observed frequency range, and the 
constant of proportionality is evaluated at the source at $r(t_e)=0$ the luminosity becomes
\be
L=\frac{{\cal L}}{4\pi R_0^2r_0^2} \frac{e^{-\tau}}{(1+z_n)^2},
\ee
where ${\cal L}$ is the absolute luminosity of an assumed isotropically radiating source in that frequency range.
The luminosity distance $d_L$ is easily read from this and differs from RW cosmology, 
\ie
\be
d_L=(1+z_n)R_0r_0\,e^{\tau/2},
\ee
 however, the apparent size distance \citep{Ellis} remains $d_A= r_0R_e$. 
These distances obviously violate the classical reciprocity relation $d_L=(1+z)^2d_A$  (see e.g. \citep{Etherington})
which is valid for non-refractive non-absorptive optics. 
For a discussion of the impact of violating the reciprocity relation on cosmology see  \citet{Bassett}.

Using the dynamics of the  FLRW cosmologies, we find
\be
d_L(z_n)=(1+z_n)\frac{c}{H_0}\frac{e^{\tau(z_n)/2}}{\sqrt{|\Omega_k|}}{\rm sinn}\sqrt{|\Omega_k|}\int_{0}^{z(z_n)}
{\frac{dz'}{n(z')h(z')}},
\label{d-z}
\ee 
where we have used Eq.\,(\ref{zn}) to express distance-redshift as a function of the frequency redshift $z_n$,
where $H_0$ is the Hubble constant, and where
\be
h(z)\equiv \sqrt{\Omega_\Lambda+\Omega_k(1+z)^2+\Omega_m(1+z)^3+\Omega_r(1+z)^4}.
\ee 
The density parameters $\Omega_\Lambda,$ $\Omega_m,$ $\Omega_r$ are standard and represent current 
relative amounts of non-interacting gravity sources: vacuum, pressureless matter, and radiation energies. 
They are related to the curvature parameter $\Omega_k$ by
\be
\Omega_k\equiv-\frac{c^2k}{H_0^2R_0^2}=1-(\Omega_\Lambda+\Omega_m+\Omega_r).
\ee 
When the optical depth is written as a function of the frequency redshift $z_n$ it becomes
\be
\tau(z_n)=\frac{c}{H_0}\int_{0}^{z(z_n)}\frac{\alpha(z')}{(1+z')n(z')h(z')}\ dz'.
\label{tau}
\ee

\begin{figure*}
\includegraphics{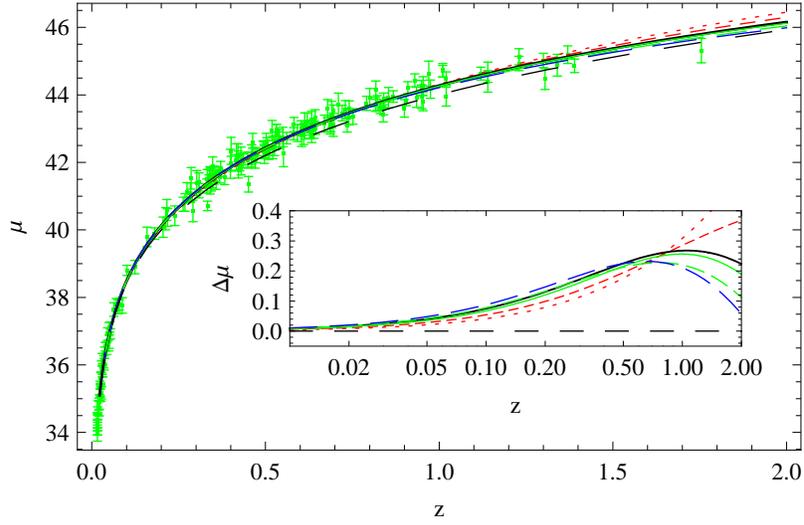}
\caption{ Distance modulus $\mu$ versus redshift $z.$ The two red and one solid black curves are open models, the two green curves are flat models, and the blue curve is a closed model.  (looking downward at redshift $z=1.5$) Dotted red curve: $\Omega_\Lambda=0,$ $\Omega_{\rm m}=0.05,$ $\alpha(z)=7\times 10^{-5}\>{\rm Mpc}^{-1}.$ Short dashed red curve : $\Omega_\Lambda=0$, $\Omega_{\rm m}=0.3,$ $\alpha(z)=1.3\times 10^{-4} \>{\rm Mpc}^{-1}.$ Solid black curve (our best fit): $\Omega_\Lambda=0$, $\Omega_{\rm m}=0.73,$ $\alpha(z)=2.2\times 10^{-4} \>{\rm Mpc}^{-1}.$ Solid green curve (concordance model): $\Omega_\Lambda=0.7,$ $\Omega_{\rm m}=0.3,$ $\alpha=0.$  Longer dashed green curve: $\Omega_\Lambda=0$, $\Omega_{\rm m}=1.0,$ $\alpha(z)=2.6\times 10^{-4} \>{\rm Mpc}^{-1}.$ Longest dashed blue curve: $\Omega_{\rm m}=1.3,$ $\alpha(z)=3.2\times 10^{-4} \>{\rm Mpc}^{-1}$. Black dashed curve: $\Omega_\Lambda=0$, $\Omega_{\rm m}=0.3,$ $\alpha=0.$ 
Inset: $\Delta\mu$ versus $z$ curve for each model, the fiducial model (black dashed curve): 
$\Omega_\Lambda=0,$ $\Omega_{\rm m}=0.3,$ $\alpha=0.$}
\label{fig:MuZPlot}\end{figure*}

\begin{figure*}
\includegraphics{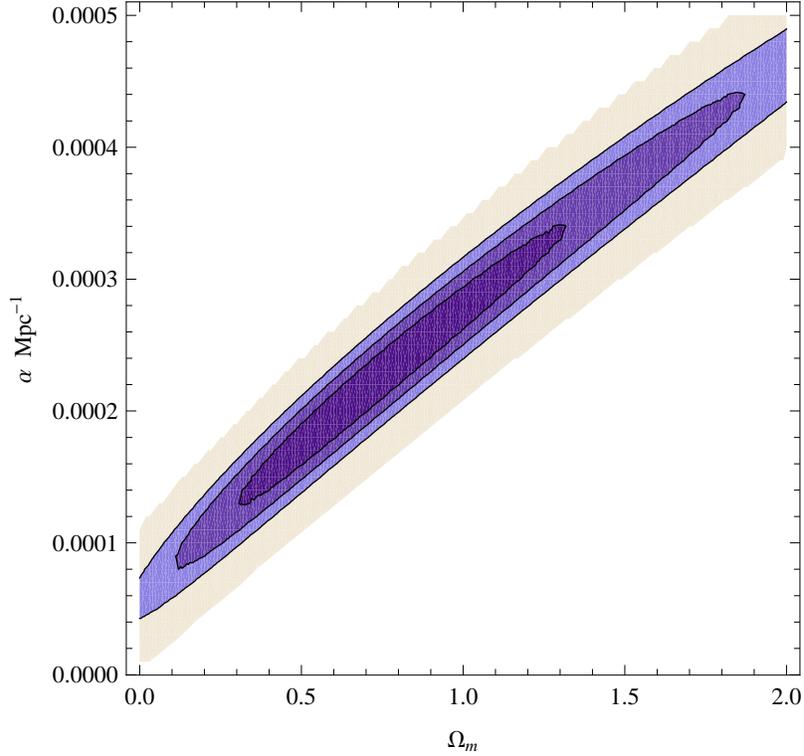}
\caption{ $\alpha$ versus $\Omega_{\rm m}.$  
The best fitting parameters are $\Omega_{\rm m}=0.73, \alpha=2.2\times 10^{-4}{\rm Mpc}^{-1},$ with  $\chi^2_{\rm min}=1.04$ (per degree of freedom). The innermost contour encloses the $68.3\%$ confident region, the next one encloses the $95.4\%$ confidence region, and the outermost one encloses the $99.73\%$ confidence region.  }\label{fig:chi2}\end{figure*}

The distance-redshift of Eq.\,(\ref{d-z}) can be compared with a similar result but without absorption given in \cite{Kantowski}. In that paper we have shown  that a cosmological model with a refraction 
index $n(z)=1+az^2+bz^3$, but with no absorption, can fit the currently available 
type Ia supernovae data quite well. A source for such refraction remains elusive. 
For illustrative purposes, we now use the distance-redshift of Eq.\,(\ref{d-z}) 
with a constant absorption coefficient $\alpha$ which does not change with frequency (\ie is grey) 
but without refraction to fit this same supernovae data, \ie we  take
\be
\Omega_\Lambda=0,\> n(z)=1,\> \alpha(z)= {\rm const}.
\ee 
The Hubble constant we use is $H_0=65\> {\rm km}\cdot {\rm s}^{-1}\cdot{\rm Mpc}^{-1}.$ Since we are concerned with the matter dominated era, we exclude radiation ($\Omega_r=0).$ The function $h(z)$ in Eqs. (\ref{d-z}) and (\ref{tau}) simplifies to
\be
h(z)=(1+z)\sqrt{1+\Omega_m z},
\ee and our model has just two parameters $(\Omega_{\rm m},\,\alpha).$
We compare the distance modulus versus redshift, $\mu(z)$, 
of  the concordance model ($\Omega_\Lambda=0.7,\>\Omega_m=0.3$) 
with five $\Omega_\Lambda=0$ extinction models (three open, one flat and one closed). 
The result is shown in Fig.\,\ref{fig:MuZPlot}, 
where the distance-modulus $\mu$, defined by
\be
\mu=5 \log \frac{d_L}{1 {\rm Mpc}}+25,
\ee  
is compared to the supernova data \citep{Riess2,Riess,Astier,Davis, Wood}.  
We use the $178$ supernova from the gold sample \citep{http}
with redshifts greater than $cz = 7000\> \hbox{km}/\hbox{s}$.  The critical 
redshift region is between 
$0.2<z<1.2,$ where most of the supernova data is concentrated.  
In Fig.\,\ref{fig:MuZPlot}, the solid green curve represents the flat concordance model dominated by dark energy. The dotted red curve is an open universe with baryonic only matter, $\Omega_\Lambda=0,$ $\Omega_{\rm m}=0.05,\ n=1$ and $\alpha(z)=7\times 10^{-5}\>{\rm Mpc}^{-1}.$ The short dashed red curve is an open model containing only dark matter,  $\Omega_\Lambda=0$, $\Omega_{\rm m}=0.3,\ n=1$ and $\alpha(z)=1.3\times 10^{-4} \>{\rm Mpc}^{-1}.$ The solid black curve is our least $\chi^2$ model (see Fig.\,\ref{fig:chi2}) with $\Omega_\Lambda=0$, $\Omega_{\rm m}=0.73,$ $\alpha(z)=2.2\times 10^{-4} \>{\rm Mpc}^{-1}.$ The longer dashed green curve represents a flat universe dominated solely by matter, with $\Omega_\Lambda=0$, $\Omega_{\rm m}=1.0,$ $\alpha(z)=2.6\times 10^{-4} \>{\rm Mpc}^{-1}.$ The longest dashed blue curve is a closed model, where $\Omega_\Lambda=0$, $\Omega_{\rm m}=1.3,$ $\alpha=3.2 \times 10^{-4} \>{\rm Mpc}^{-1}.$ The black dashed curve is the now disfavored matter dominated  
$\Omega_\Lambda=0$, $\Omega_{\rm m}=0.3$ model (without extinction). 
In the inset we show  $\Delta\mu$ versus  $z$ (relative to the disfavored matter dominated case) 
for each model. To produce roughly the same amount of change in the distance-modulus 
a larger absorption coefficient $\alpha$ is needed for a larger $\Omega_{\rm m}$ ($\alpha$ is positively correlated to $\Omega_{\rm m}$). As the reader can easily see 
in Fig.\,\ref{fig:MuZPlot} the 
effects of a suitable value of the simplest absorption coefficient $\alpha(z)={\rm const}$ can 
simulate the accelerating effects of a cosmological constant.

In Fig.\,\ref{fig:chi2}, we show the confidence contours for our model parameters, $\alpha$ versus $\Omega_{\rm m}$.  The best fitting parameters are $\Omega_m=0.73, \alpha=2.2\times 10^{-4}{\rm Mpc}^{-1},$ with $\chi^2_{\rm min}=1.04$ (per degree of freedom). The innermost contour encloses the $68.3\%$ confidence region, and the next one  encloses the $95.4\%$ confidence region, the outermost one encloses the $99.73\%$ confidence region. For each fixed $\Omega_{\rm m}$ selected in Fig.\,\ref{fig:MuZPlot}, $\alpha$ is the value that gives the least $\chi^2.$

If we extract a density factor $\rho$ from the absorption coefficient, 
\ie $\sigma\equiv \alpha/\rho$ we obtain an opacity in \eg ${\rm cm}^2\cdot {\rm g}^{-1}$.
The density $\rho$ with units of ${\rm g}\cdot{\rm cm}^{-3}$ is 
the density of the relevant species causing the absorption. 
A competitive absorption model should properly account for both 
the cosmic expansion and the physical/chemical evolution of the
inter-galactic medium \citep{Spitzer}.
Here we are content with an order of magnitude estimate noting 
that  an  opacity $\sigma=10^5\, {\rm cm}^2\cdot{\rm g}^{-1}$ 
as proposed for the carbon needle model in \citet{Aguirre}, requires a density $\rho$ the order of $10^{-33}\, {\rm g}\cdot {\rm cm}^{-3}$ 
to produce the absorption needed 
for the above fitting. This density is only a factor of $\sim 10^{-4}$ 
of the current critical mass density 
$\rho_c\approx 8\times 10^{-30} {\rm g}\cdot{\rm cm}^{-3}.$    
For fine-tuned dust absorption models, 
see e.g., \citep{Aguirre,Aguirre2, Goobar,Bianchi, Robaina}.


\eject
\section{Discussion}\label{sec:Discussion}

We have demonstrated how the 4-D optical metric of Gordon \citep{Gordon} [see Eq.\,(\ref{Gordon-Metric})] 
can be extended 
and used even in cases where absorption is present. We looked at both `strong' and `weak' 
absorption, the distinction  being whether absorption is significant on wavelength scales or 
only over a multitude of wavelengths. The two cases are distinguished respectively by complex
and real eikonals. For the complex eikonal case the optical metric must remain complex 
(see Sec.\,\ref{sec:Complex}), however, for the 
weak absorption case the real part of the optical metric (essentially the same as 
Gordon's original proposal) remains as the significant geometrical structure (see Sec.\,\ref{sec:Real}).  
The two cases 
differ on how absorption appears in the geometrical optics field. In the `strong' absorption case the 
imaginary part of the eikonal reduces the wave's intensity [see Eq.\,(\ref{4S})] but in the `weak' case the amplitude's 
reduction [see Eq.\,(\ref{tilde-Eprop})] is responsible for  the intensity decrease. 

A geometrical optics wave is like a single frequency wave even though the wave's frequency changes from 
spacetime point to spacetime point. To superimpose multiple frequencies is straightforward,
however, to superimpose optical metrics, real or complex, makes no sense. Consequently a
useful single optical metric only exists when the optical properties are insensitive 
to the superimposed frequencies. Such frequency independence approximations are often designated as `grey' 
in astrophysical  applications.

The two examples we gave in Section \ref{sec:Complex} can be used to study the impact of refraction and/or absorption on light propagation in stellar atmospheres. The classical radiation transport equation (see e.g. \cite{Lindquist, Ehlers3}) is derived assuming that light follows null geodesics in curved spacetimes. With the presence of light refraction, both the direction and speed of light change. This would require the radiative transfer equation to be written using the optical metric instead of the physical metric. The optical metric might also be of some use in hydrodynamical simulations of stellar interiors. For an example, the slowing down of light will reduce the efficiency of energy transport outward via the radiation field. Our first example (flat spacetime) can be used to study the 1-D case; a comparison of results of a numerically solved radiation transport equation with/without a refraction index $n(z)$ would be interesting. Similarly, our second example (curved spherically symmetrical spacetime) can be used to model the atmospheres of neutron stars. Further non-spherically symmetric examples could be useful in radiation transport calculations in accretion disks of black holes. 

Our last example, the real eikonal case, allowed us to give luminosity distance redshift for 
observations in standard cosmologies where both refraction and absorption were present. 
As an example of the usefulness of this theory 
we went on to fit the gold sample of type Ia supernovae to a Hubble curve corrected for grey extinction. 
In Section \ref{sec:Real} we showed that it is possible to explain the current supernova observations via 
a simple absorption model instead of requiring the existence of dark energy. Our best fit was an 
open $\Omega_m=0.73$ model with constant absorption $\alpha=2.2\times 10^{-4}{\rm Mpc}^{-1}$, 
see Fig.\,\ref{fig:chi2} for confidence contours. More realistic $z$ dependent models for $\alpha$ are in order.
Since the flat concordance model is supported by other observations, e.g., 
cosmic microwave background data and baryonic acoustic oscillations, an absorption theory cannot be 
on firm ground unless it provides an explanation for these additional observations.  
We leave these and other applications to future efforts.

\section{Acknowledgments}

This work was supported in part by NSF grant AST-0707704 and US DOE
Grant DE-FG02-07ER41517. The authors wish to pay tribute to the memory of Jürgen Ehlers for the
enlightenment and encouragement he gave one of us many years ago.

\label{lastpage}

\end{document}